\documentstyle[twocolumn,prl,aps,epsf]{revtex}

\newcommand{\epsplace}[1]{\epsfbox{#1}}

\begin{document}


\title{Freezing Transition of Compact Polyampholytes}
\author{ Vijay S. Pande\dag, Alexander Yu. Grosberg\ddag, Chris Joerg\S,
Mehran Kardar, and Toyoichi Tanaka}
\address{ Department of Physics and Center for Materials Science and
Engineering, \\ Massachusetts Institute of Technology, Cambridge,
Massachusetts 02139,  USA \\
\dag {\em Present address:\/} Physics Department, University of
California, Berkeley, California 94720-7300, USA\\
\ddag {\em On leave from:\/} Institute of Chemical Physics,  Russian
Academy of Sciences, Moscow 117977, Russia\\
\S Laboratory for Computer Science, Massachusetts Institute of
Technology, Cambridge, Massachusetts 02139,  USA }

\address{ {\em \bigskip
\begin{quote}
Polyampholytes (PAs) are heteropolymers with long range Coulomb
interactions. Unlike polymers with short range forces, PA
energy levels have non-vanishing  correlations and 
are thus very different from the Random Energy Model
(REM). Nevertheless, if charges in the PA globule are screened as in
a regular plasma, PAs freeze in REM fashion.
Our results shed light on the potential role of Coulomb interactions  
in 
folding and evolution of {\it proteins}, which are weakly charged  
PAs,
in particular making connection with the finding that 
sequences of charged amino acids in proteins are not random.
\end{quote} } }


\maketitle



The freezing transition of heteropolymers, in which the number of
thermodynamically relevant states goes from an exponentially large
value (${\cal O}(e^{N})$) in the random globule state, to  only a
few (${\cal O}(1)$) conformations in the frozen state,  has
attracted a great deal of interest. In addition to providing an
interesting problem in the statistical mechanics of disordered
materials\cite{Previous}, this system is
potentially relevant to the biologically important question of
protein  folding.
Most previous investigations have focused on  heteropolymers
with short-range interactions.  Recently, however, there has
been renewed theoretical\cite{Polyamph,KLK,DobRub} and
experimental\cite{Copart,Tanaka} interest in polyampholytes (PAs),
which are heteropolymers with charged monomers of both signs.
It has been shown that, due to screening effects, PAs
collapse to compact globules if their net charge is below a
critical value\cite{KK}.  There is also some evidence from exact
enumeration studies of short chains\cite{KKenum} that
dense globules  of neutral PAs may have a freezing transition.
However, it is unclear how long range (LR) interactions
affect  freezing, or  whether the formalism developed for globular
polymers with short range (SR) interactions remains
applicable to the LR case.

The freezing transition of SR heteropolymers is most commonly
described by the Random Energy Model (REM)\cite{Derrida},
although it is not always applicable even in this
case\cite{HowGoodREM}).
As the principle underlying assumption of REM is the statistical
independence of energies of states (polymer conformations) over
disorder (sequence of charges along the chain), we first examine
correlation of the energies  and then discuss the
resulting freezing transition.
Our starting point is the Hamiltonian
\begin{equation} {\cal H} = \sum_{I \neq J}^N B s_I s_J f({\bf
r}_I  - {\bf r}_J) ,
\end{equation}
where $B$ is a constant, $I$ labels monomers along the chain,
and $s(I) \in \pm 1$ is the charge of monomer $I$. The range of
interactions is indicated through $f(r)$, such that $f(r) = \Delta (r)$
for SR interactions, and $f(r)=1/r^{d-2}$ for Coulomb forces in $d$
dimensional space.
Finally, we only consider the case of
maximally compact polymers, assuming that maximal density is
maintained independently of Coulomb interactions, i.e. by an external
box, poor solvent, or internal attractions, such that $R \sim  N^{1/d}$.

The simplest characteristics of statistical dependence of energies
is the pair correlation between two arbitrary conformations
$\alpha$ and $\beta$, given by
\begin{equation}
\left<  E_{\alpha}E_{\beta} \right>_c \equiv
\left< E_\alpha E_\beta \right> - \left< E_\alpha \right>
\left<E_\beta \right>
= B^2 {\cal Q}_{\alpha \beta} \ ,
\label{eq:E1E2}
\end{equation}
with ${\cal Q}_{\alpha \beta} \equiv
\sum_{I \neq J} f({\bf r}_I^\alpha  - {\bf r}_J^\alpha) f({\bf
r}_I^\beta  - {\bf r}_J^\beta)$.
In the familiar case of SR interactions, ${\cal Q}_{\alpha
\beta}^{\rm SR} = \sum_{I \neq J}
\Delta({\bf r}_I^\alpha  - {\bf r}_J^\alpha)
\Delta({\bf r}_I^\beta  - {\bf r}_J^\beta)$
is just the number of bonds in common between configurations
$\alpha$ and $\beta$.
Numerical simulations\cite{HowGoodREM} indicate that in many
cases the probability distribution for ${\cal Q}_{\alpha
\beta}^{\rm SR}$, i.e. $P_{\rm SR}({\cal Q}) \equiv \sum_{\alpha
\beta} \delta({\cal Q} - {\cal Q}^{\rm SR}_{\alpha \beta})$ is sharply
peaked at small ${\cal Q}$.  This happens because one can easily
``hide'' monomers by moving them only a small distance and
decreasing their contribution to ${\cal Q}^{\rm SR}$.  Large
statistical dependence is thus achieved only for conformations
that are closely related.   The validity of REM rests on the statistical
rarity of such closely related conformations. REM is valid when
configurations that are statistically dependent can be ignored in
a large $N$ limit.

By contrast, with long range interactions, the relevant
parameter for judging statistical dependence is
${\cal Q}_{\alpha \beta}^{\rm LR} = \sum_{I \neq J}  [|{\bf
r}_I^\alpha-{\bf r}_J^\alpha| \cdot |{\bf r}_I^\beta-{\bf
r}_J^\beta|]^{-(d-2)}$. While the geometric interpretation of
${\cal Q}_{\alpha \beta}^{\rm LR}$ is not as clear as ${\cal
Q}_{\alpha \beta}^{\rm SR}$, it measures the similarity in
contributions from monomer pairs $(I,J)$ in conformations
$\alpha$ and $\beta$ to the overall energy.
Unlike the SR case,   polymeric bonds always keep monomers
within the scale of LR interactions.  Thus, for two conformations
chosen at random, the overlap ${\cal Q}^{\rm LR}_{\rm rand}$
may not be negligible (even if ${\cal Q}^{\rm SR}_{\rm rand}$ is).
The following scaling argument provides an estimate of the width of
the probability distribution $P_{\rm LR}({\cal Q}) \equiv
\sum_{\alpha \beta} \delta({\cal Q} - {\cal Q}^{\rm LR}_{\alpha
\beta})$.

\begin{figure}
\epsfxsize=3.3in \centerline{ \epsplace{RandScalingQvsN.eps.art} }
\caption{
Scaling of ${\cal Q}_{\rm rand}$  and
${\cal Q}_{\rm max}$  with $N$ for LR and SR interactions ($d=3$).
Power law scaling of the form  ${\cal Q} \sim N^\gamma$ indicates
that
${\cal Q}^{\rm LR}_{\rm rand}/{\cal Q}^{\rm LR}_{\rm max}$ does not
vanish in the thermodynamic limit, whereas
${\cal Q}^{\rm SR}_{\rm rand}/{\cal Q}^{\rm SR}_{\rm max}$ does.
}\end{figure}


\begin{figure}
\epsfxsize=3.3in \centerline{\epsplace{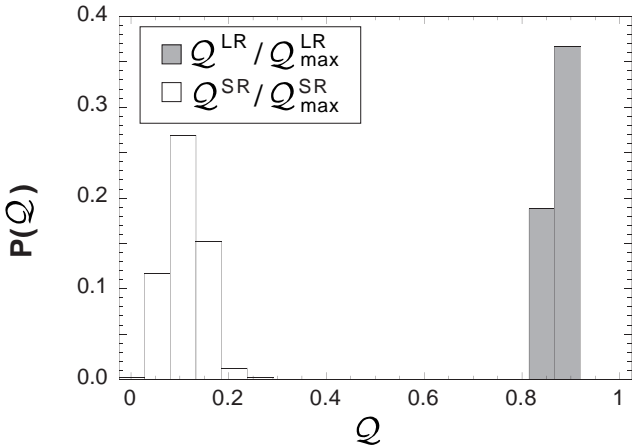}}
\caption{
Probability distributions $P({\cal Q}^{\rm LR})$  and $P({\cal Q}^{\rm
SR})$,
obtained from 64-mers on a cubic lattice.  Due to finite size effects,
there is
some
residual overlap in the SR case (here peaked at 0.1).  However, we
expect that the SR residual overlap vanishes in the thermodynamic
limit, while the LR overlap does not.
}\end{figure}


First,  consider the maximum overlap which occurs (for both LR
and SR) when {\it all} elements are correlated (i.e. ${\cal Q}_{\rm
max}=
{\cal Q}_{\alpha \alpha}$ is the correlation of a configuration with
itself).
To compute this, we note that for each of the $N$ monomers, there is a
contribution from ${\cal O}(r^{d-1})$ monomers at a distance $r$
(for {\it compact} states in $d$ dimensions), resulting in
${\cal Q}_{\rm max} \sim N \int dr  r^{d-1} f(r)^{2}$.  For SR
interactions, this integral is dominated by contributions at a
microscopic length scale (set by the interaction range)
and we get  ${\cal Q}^{\rm SR}_{\rm max} \sim N $.  For LR
interactions, while contributions from monomers far
away are smaller, there are more of them.  For Coulomb interactions
in $d\leq 4$, the integral is dominated by the longest distance, and
for a polymer of size $R$, we get
${\cal Q}^{\rm LR}_{\rm max} \sim N R^d/R^{2(d-2)} \sim
NR^{4-d}$.

We can use similar arguments for the
overlap between two conformations chosen at random 
(${\cal Q}^{\rm LR}_{\rm rand}$).
In fact, for the LR problem,
 ${\cal Q}^{\rm LR}_{\rm max}$ and ${\cal Q}^{\rm LR}_{\rm rand}$ scale
identically,
as both cases involve ${\cal O}(N^2)$ pairs of monomers
each giving a  contribution ${\cal O}(1/R^{2(d-2)})$, for a
total of ${\cal Q}^{\rm LR}_{\rm max} \sim
{\cal Q}^{\rm LR}_{\rm rand} \sim N^{2}R^{2(2-d)}$.  
 Moreover, as
the main contribution to ${\cal Q}^{\rm LR}_{\rm rand}$ comes from
far away sites, this residual overlap is only weakly conformation
dependent.
 The existence of a
residual overlap changes the problem fundamentally from the
SR case:  REM is not valid as there is always a  statistical
dependence in $d < 4$ \cite{Note1}.

Computer simulations support the above arguments.  To examine a large
range in $N$, we generated random conformations on a lattice by
first choosing a radius $R$, and then enumerating random paths
\cite{Enum}
on the set of lattice sites which are within $R$.
$R$ was varied from 3 to 10 lattice sites, and the following results
represent averages over 20 conformations for each $R$ value.
Fig.~1 shows  that the scaling exponents $\gamma$  defined by
${\cal Q} \sim N^\gamma$ appear to be the same within error for random
pairs of conformations, as well as the overlap of any conformation with
itself. Furthermore, the fits agree well with the predictions
$\gamma^{\rm LR}_{\rm max} = \gamma^{\rm LR}_{\rm rand} = 4/3$.
By contrast, with SR interactions $\gamma^{\rm SR}_{\rm max} = 1$,
while $\gamma^{\rm SR}_{\rm rand} \approx 0.75$ is distinctly smaller.
We also calculated SR and LR overlaps ${\cal Q}^{\rm SR}$ and
${\cal Q}^{\rm LR}$ for 1000 pairs of 64-mer conformations ($d=3$,
cubic lattice).  The resulting histograms, with overlaps normalized
by the maximal value, are shown in Fig.~2.  SR overlaps are peaked
at small values whereas the LR overlaps are peaked closer to
unity.   Furthermore, the sharpness of the distribution suggests that
${\cal Q}^{\rm LR}$ is approximately independent of the chosen
pairs of conformations.

Having demonstrated the residual overlap between energies
of conformations with LR interactions, and hence the breakdown
of REM, we go on to better characterize the density of states.
This will take us a step closer to understanding the freezing of PAs.
To describe the density of states, we use the following three
characteristics:  the annealed energy variance $\sigma_{\rm ann}$ (the
width of the density of states for annealed disorder), the average
quenched
energy variance $\sigma_{\rm quen}$ (the width of the density of states
for quenched disorder), and  the quenched energy correlation function
$g$ (the statistical dependence between states). These quantities are
given by the formul\ae

\begin{eqnarray}\label{define}
\sigma^2_{\rm ann} & \equiv & \left< \overline{(E^2)} \right>_c =
\left< \overline{(E^2)} \right> - \left<\, \overline{E}\, \right>^2,
\nonumber \\
\sigma^2_{\rm quen} & \equiv &\left< \overline{(E^2)}_c \right> =
\left< \overline{(E^2)} \right> - \left< (\overline{E})^2
\right>,
\\ g & \equiv & \left< {(\overline{E} )}^2 \right>_c  = \left<
{(\overline{E} )}^2 \right> -
\left<\, \overline{E} \,\right>^2,
\nonumber
\end{eqnarray}
where $\overline{ {}^{ } \ldots }$ and $\left< \ldots
\right>$ denote averaging over conformations and sequences
respectively.  Note that these quantities are related by a
mathematical identity
$\sigma^2_{\rm ann} = \sigma^2_{\rm quen} + g$.

\begin{figure}
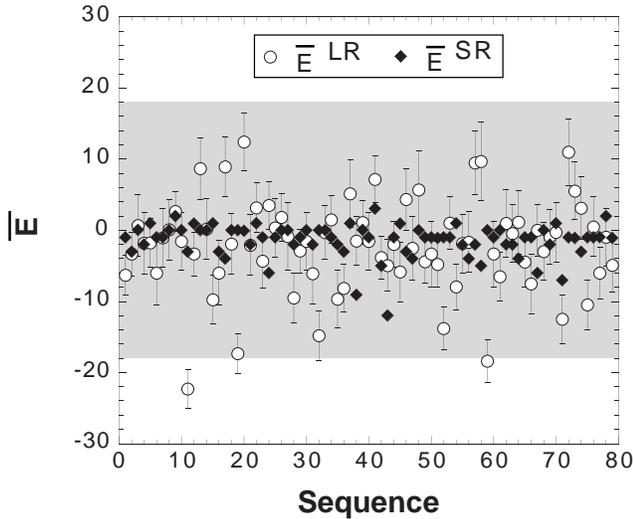

\epsfxsize=3.3in \centerline{ \epsplace{meanE.eps.art} }
\caption{
Mean and width of the energy spectra  for
80 sequences of 36-mers, determined by full enumeration over all
maximally compact conformations (see text for details).
}\end{figure}
\bigskip


In the annealed case, the energy variance is $\sigma^2_{\rm ann} =
B^2 {\cal Q}_{\rm max}$, since, in this case, all possible states can be
accessed and thus the width of the energy spectrum must be  maximal.
This result is also easily extracted from equation (\ref{eq:E1E2}) by
averaging over conformations with $\alpha = \beta$.   Averaging the same
equation over {\it all pairs} of  states $\alpha$ and $\beta$, we can
find $g$:
for ${\cal M}$ conformations, there are ${\cal M}$ pairs $\alpha=\beta$
which completely overlap  ${\cal Q}_{\alpha \beta} = {\cal Q}_{\rm
max}$,
but this is overshadowed by the remaining ${\cal M}({\cal M}-1)$ pairs
with overlap ${\cal Q}_{\alpha \beta}= {\cal Q}_{\rm rand}$, resulting
in $g \approx B^2 {\cal Q}_{\rm rand}$.
In addition to measuring the statistical dependence between states,
$g=\left< (\overline{ E })^2 \right>_c$ also describes how the mean of
the
energy spectrum for a given sequence varies between sequences.
Finally the width of the energy spectrum for a typical sequence is
$\sigma^2_{\rm quen} \equiv \sigma^2_{\rm ann}-g = B^2 ({\cal
Q}_{\rm max} - {\cal Q}_{\rm rand})$. This makes sense physically
as correlation (anticorrelation) in the energies should narrow
(broaden) the width of the energy spectra.  Also, we see that when
there is no correlation ($g=0$),
$\sigma_{\rm ann} = \sigma_{\rm quen}$, as in the REM.

The following picture emerges from the above results.
As ${\cal Q}^{\rm SR}_{\rm rand}=0$, we have $g=0$ for
the SR case above the  freezing temperature, and the
mean of the energy spectrum does not vary significantly
between sequences.  Also, the width of the spectrum for a
given sequence is large (the maximum possible value, as
in the annealed case).  The variation of the means of the energy
spectra between sequences $g$, is much smaller
than the typical width of each spectrum $\sigma^2_{\rm quen}$;
thus disorder is not important  for SR interactions above freezing.
Of course, below the freezing temperature, self averaging breaks
down, and disorder is relevant. By contrast, for  LR interactions,
${\cal Q}^{\rm LR}_{\rm rand}$ does not vanish and is significant.
We thus expect the widths of the energy spectra to be small
and the means to vary widely from sequence to sequence.

The results of a computational test of the above scenario,
obtained from the exact enumeration of all globular states of
36-mers on a cubic lattice ($d=3$) are presented in Fig.~3.
We see that for SR interactions, the means of the spectra are
indeed well defined and their width (gray region) is large.
For LR interactions, the means are poorly defined, with a
variance between sequences which is greater than the
widths of individual spectra (error bars).


Is the insight gained above sufficient to analyze the freezing
transition in
PAs?
In general, freezing is governed by the low energy tail of the density
of
states $\rho (E) = {\cal M} P(E)$, where ${\cal M}$ is the total number
of
conformations, and $P(E)$ is the single level energy distribution.
In the standard REM entropy crisis scenario, the system freezes in a
microstate, much like a snapshot, at a
temperature $T_f$ at which $\rho _T \sim 1$, where $\rho _T =\rho(E_T)$
is the density of states at the equilibrium energy $E_T$ at the
temperature
$T$.

The density of states in the high temperature regime is governed by
$\sigma _{\rm ann}$, as can be seen by a high temperature expansion:
The partition function $Z={\rm tr}\left[ \exp\left( -\beta{\cal H}
\right)
\right]$
is first expanded in powers of $\beta=1/T$, resulting in (after
averaging over
sequences) $-\beta F=\left\langle\ln Z\right\rangle=\ln{\cal M}-
\beta\left\langle \overline{E} \right\rangle+
\beta^2\langle \overline{(E^2)} \rangle_c/2+\cdots$.
{}From this expression (and using Eq.(\ref{define})), the entropy is
calculated as $S(T)=\ln {\cal M}-\beta^2\sigma^2_{\rm quen}/2+\cdots$,
where (as demonstrated earlier) for Coulomb interactions in $d=3$,
$\sigma^2_{\rm quen}\sim e^2N^2/R$, 
 yielding  
\begin{equation}
\rho_T \sim
{\cal M} 
\exp \left[- {1\over 2}\left({e^2 N \over TR} \right)^2 \right] .
\end{equation}

>From the structure of the series\cite{KLK}, we expect the high
temperature
expansion to break down for temperatures $T<T_D \equiv e^2N/R$. This
temperature can also be obtained by regarding the polymer globule as a
(non-polymeric) plasma of the same $N$ charges confined within the
volume
$R^3$. As the Debye screening length for this plasma is of the order
$r_D \sim (TR^3/N e^2)^{1/2}$, there are two regimes: For $T<T_D$, the
plasma
is fully screened as $r_D < R$. However, for $T>T_D$, $r_D > R$ and the
charges are not screened. The latter regime is meaningless for a regular
plasma, but describes the high temperature behavior of the polymer
globule. 
It is not clear that, with the constraints of polymeric bonds, the
scaling
for a PA should be the same as that for a screened plasma at low
temperatures.   However, assuming that this is the case, the entropy can
be
estimated by noting that the plasma is composed of roughly
${\cal N} \sim R^3/r_D^3\sim (Ne^2/RT)^{3/2}$ independent Debye volumes.
Assuming that the entropy is proportional to ${\cal N}$, we finally
conclude
\begin{equation}
\rho_T \sim
{\cal M} 
\exp \left[-c \left( \frac{e^2 N}{TR} \right)^{3/2} \right]
\label{eq:renormdos}
\end{equation}
where $c$ is a numerical constant.  Note that
Eq.~(\ref{eq:renormdos}) indicates a very sharp decrease of the density
of
states in the low energy tail, proportional to $\exp [ - c^{\prime} ( E
-
\overline{E} )^3 ]$, which reflects the fine tuning of configurations
necessary
for screening.

Typically the number of conformations of a polymer scales as
${\cal M} \sim e^{\omega N}$, with $\omega$ of the order of unity.
In the limit where the polymer is kept maximally compact by an external
box, poor solvent, or internal attractions, such that $R \sim a
N^{1/3}$,
where $a$ is a monomeric length scale, $\omega$ is approximately the
entropy of Hamiltonian walks.
Freezing, which is signaled by $\rho \sim 1$,
can take place in the unscreened regime only for short chains with
$N < 1/\omega$.  (The ``apparent'' freezing temperature for unscreened
polymers grows as $N^{1/6}$.) In this case, a further decrease of
temperature will not lead to screening, of course.  For longer chains,
we predict freezing at an $N$-independent temperature of
$T_f\sim e^2/(a\omega^{2/3})$ in the screened regime. In this sense, the
compact PA freezes in a phase transition that is similar to REM.  We
stress that this happens despite the unusual scaling of the width of the
density of states, $\sigma \sim N^{2/3}$. The distinction between the
two
behaviors is important for understanding the results of lattice
simulations, as it appears that 36-mers are in the short chain regime.

We expect that the nature of the frozen state also depends on $T_f/T_D$.
For freezing in the screened regime ($T_f<T_D$), the system looks much
like
that of the SR case, i.e. like a disordered version of a salt crystal.
For freezing in the unscreened regime  ($T_f>T_D$), we expect a smaller
degree of antiferrogamnetic ordering; consistent with the
idea
that freezing at a higher temperature leads to a state which is less
energetically optimized.

An important class of PAs are {\it proteins}.  In the light of our
findings 
in this work, we make here some concluding remarks about protein
folding and evolution. 
Of the 20 natural amino-acids, three are positively charged (Lys, Arg,
His),
two are negatively charged (Asp, Glu), and the rest are neutral. 
Nevertheless, it is often assumed that LR interactions are not
essential to proteins, as the screening length in biological solvents is
often quite small. It is less clear that screening is also effective in
compact globular configurations with little or no solvent in their
interiors. Furthermore, secondary structural elements such as
$\alpha$-helices effectively reduce the conformational flexibility of
proteins. Indeed, the conformation space of small proteins (i.e. 70-90
amino--acids) perhaps corresponds to that of lattice 27-mers
\cite{CoresStates},
and small proteins are likely to be in the short chain regime with
respect to LR interactions. Thus, while the total charge on a given
protein
may be small, in solvents with few counter ions, this may be sufficient
to
lead to a REM-violating correlated energy landscape, making the
results
obtained here  relevant.  Moreover, for the
typical separation of charges in a globular protein (roughly  
20 \AA), and given a dielectric constant of order 5-10, and  
$\omega\approx 2$, the characteristic freezing temperature $T_f$ is of  
the order of (biologically relevant) room temperatures.

We have discussed how the mean of the density of states can vary greatly
from
sequence to sequence.  It appears that a large contribution to this mean
comes from the interaction between monomers that are not far apart along
the
sequence. For example, while next nearest neighbors along the chain can
somewhat vary their spatial distance from each other, this will still
not
break their great contribution to the mean energy.  This is why the
conformational average energy depends strongly on the correlations
between
charges quenched along the sequence.   For Coulomb interactions,
chains
with anti-correlated sequences have low mean energies.  This is
intriguing, considering the recent finding that protein sequences are
indeed
anti-correlated with respect to their charge \cite{ProtCor}.   This
indicates
that perhaps protein evolution was not just dictated solely by the
degree of
hydrophobicity of monomers (which depends on the degree of charge, not
the
sign), but by Coulomb effects as well.

\bigskip

The work was supported by NSF (DMR 94-00334).
AYG acknowledges the support of Kao Fellowship.  Computations were
performed on Project SCOUT (ARPA contract MDA972-92-J-1032).

\end{document}